\documentstyle[sprocl]{article}

\def\be{\begin{equation}}
\def\ee{\end{equation}}
\def\beq{\begin{equation}}
\def\Mpl{M_{{\rm Pl}}}
\def\eeq{\end{equation}}
\def\bea{\begin{eqnarray}}
\def\eea{\end{eqnarray}}

\begin{document}

\title{D-TERM INFLATION: THE GOOD, THE BAD AND THE UGLY\footnote{
Based on the invited plenary talk given at the 
{\it COSMO97} Conference, Ambleside, England, September 15-19 1997 and on the
talk given at the {\it Phenomenological Aspects of Superstring
Theories} (PAST97) Conferences, ICTP Trieste, Italy, October 2-4 1997. Preprint
OUTP-97-55-P.}}

\author{A. RIOTTO }

\address{Theoretical Physics Department, University of Oxford,\\
1 Keble Road, OX1 3NP, Oxford, UK.}


\vskip 1cm

\maketitle\abstracts{An inflationary 
stage dominated by a $D$-term avoids 
the slow-roll problem of inflation 
in supergravity and can naturally emerge 
in theories with a non-anomalous or anomalous 
$U(1)$ gauge symmetry. In this talk advantages and disavantages
 of
$D$-term inflation are  discussed.}

{\it Prologo.}~~It is by now commonly accepted that  
inflation   looks  more natural in supersymmetric theories
 rather in non-supersymmetric ones. This is because  
the necessity of introducing very small
parameters to ensure the extreme flatness of the inflaton potential 
seems very
unnatural and fine-tuned in  most non-supersymmetric theories, while 
this naturalness is achieved in supersymmetric models. The nonrenormalization
 theorems in exact global supersymmetry guarantee 
that we can fine-tune any parameter at the tree-level and this 
fine-tuning will not be destabilized by radiative corrections at any order
 in perturbation theory. This is the advantage of
 invoking supersymmetry. There is, however, a severe problem one has to 
face when dealing with inflation model building in the context of 
supersymmetric theories. The  generalization of supersymmetry
 from a global to 
a local symmetry automatically incorporates gravity and, 
therefore, inflation model building must be considered
 in the framework of supergravity theories. 
The supergravity potential is rather involved, but it  can still  be written as 
a $D$-term plus an $F$-term, and it is usually supposed that
the $D$-term vanishes during inflation. 
Now, for models where the $D$-term vanishes,
the slow-roll  parameter $\eta=\Mpl^2 V''/V$ generically receives
various contributions of order $\pm 1$. 
This is the so-called $\eta$-problem of supergravity 
theories\cite{CLLSW,coughlan}.
In supergravity theories, supersymmetry breaking is transmitted 
by gravity interactions and the squared mass of the inflaton 
becomes naturally of order of $V/\Mpl^2\sim  H^2$. 
The perturbative renormalization of the K\"ahler 
potential is therefore crucial   for the 
inflationary dynamics due to a non-zero energy
density which breaks supersymmetry spontaneously 
during inflation. How severe the problem is depends on the magnitude of $\eta$
and it is crucial to notice that the amount of fine-tuning needed is
directly indicated  by the present data on the cosmic microwave background
anisotropy. Indeed,  having $\eta$ not too small requires 
that the spectral index  $n=1-6\epsilon+2\eta$ 
($\epsilon=\frac{1}{2}\Mpl^2(V'/V)^2$ is another slow-roll parameter) 
be not too small, so the observational bound
$|n-1|<0.3$ is already beginning to make an accident look quite unlikely.
Several proposals to solve   the $\eta$-problem already
 exist in the  literature and we refer the 
interested reader to the long review \cite{lr} for a complete list and
discussion.
In this talk, we will restrict ourselves to what we believe is  the most 
promising solution: $D$-term inflation \cite{bindvali,halyo,dr}. 
$D$-term inflation   is based on the observation that $\eta$ gets
 contributions of order one  only if inflation proceeds 
along a $D$-flat direction or, in other words, 
when the vacuum energy density is dominated by an $F$-term. 
On the contrary, if  the  vacuum energy density is dominated by nonzero
$D$-terms  and supersymmetry breaking is of the $D$-type, 
scalars get supersymmetry soft breaking masses 
which depend only on their gauge charges. Scalars charged under  
 the corresponding gauge symmetry obtain a mass much larger than $H$, 
while gauge singlet fields can only get
 masses from loop  gauge interactions.  In particular, 
if the inflaton field is identified with a gauge singlet, 
its potential may be flat up to loop corrections and 
supergravity corrections to $\eta$ from the $F$-terms are not present. 

{\it  How it works.}~~
If the theory  contains an abelian  $U(1)$  gauge symmetry (anomalous or not), 
the Fayet-Iliopoulos
$D$-term term $
\xi \int d^4 \theta ~V = \xi D$
is gauge invariant and therefore allowed by the symmetries.  
It may lead  to $D$-type  supersymmetry breaking.
 It is important to notice that {\it 1)} 
this term may be present in 
the underlying theory from the very beginning; {\it 2)} it may  appear in the
 effective theory after some heavy degrees of freedom have been integrated out;
{\it 3)} it looks  particularly intriguing 
that an  anomalous $U(1)$
symmetry  is  usually present in string theories \cite{u(1)A}.  
 The corresponding Fayet-Iliopoulos term is  \cite{fi} $
\xi = \frac{g^2}{192\pi^2}\:{\rm Tr} {\bf Q}\:\Mpl^2$, 
 where ${\rm Tr} {\bf Q}\neq 0$ indicates 
the trace over the $U(1)$ charges of the fields present in 
the spectrum of the theory.
The $U(1)$ group may be assumed to emerge from string theories so that 
the  anomaly is cancelled by the Green-Schwarz
mechanism. In such a case $\sqrt{\xi}$ is expected to be of the order of the 
stringy scale,   $(10^{17}-10^{18})$ GeV or so.
Let us briefly remind the reader how $D$-term inflation 
proceeds \cite{bindvali,halyo}. To exemplify the description, 
let us take the  toy model containing  three chiral 
superfields $S$, $\Phi_+$ and
$\Phi_-$ with charges equal to $0$, $+ 1$ and $- 1$
 respectively under the $U(1)$ gauge symmetry.
The superpotential has the form $
W = \lambda S\Phi_+\Phi_-$.
The scalar potential in the global supersymmetry limit reads
\begin{equation}
V = \lambda^2 |S|^2 \left(|\phi_-|^2 + |\phi_+|^2 \right) +
\lambda^2|\phi_+\phi_-|^2 + 
{g^2 \over 2} \left(|\phi_+|^2 - |\phi_-|^2  + \xi \right)^2
\end{equation}
where $\phi_{\pm}$ are the scalr fields of the supermultplets $\Phi_{\pm}$, 
 $g$ is the gauge coupling and $\xi>0$ is a Fayet-Iliopoulos $D$-term. 
The global minimum is supersymmetry conserving, but the gauge group
 $U(1)$ is spontaneously broken, $
\langle S \rangle  = \langle \phi_+  \rangle = 0, ~~~ \langle \phi_-\rangle 
 = \sqrt{\xi}$.
However, if we minimize the potential, for  fixed values of $S$, with respect to
other fields, we find that for  $S > S_c = {g \over \lambda}
\sqrt{\xi}$, the minimum is at $\phi_+ =\phi_- = 0$. Thus, for
$S > S_c$ and $\phi_+ =\phi_- = 0$ the tree level potential
has a vanishing curvature in the $S$ direction and large positive
curvature in the remaining two directions $
m_{\pm}^2 = \lambda^2|S|^2 \pm g^2\xi$
For arbitrarily large $S$ the tree level value of the potential remains
constant $V = {g^2 \over 2}\xi^2$ and the  $S$ plays 
the role of the   inflaton. As stated above, the charged fields get very 
large masses due to the $D$-term supersymmetry breaking,
 whereas the gauge singlet field is massless at the tree-level. 
What is crucial is that,  along  the inflationary
trajectory $\phi_{\pm}=0$, $S\gg S_c$, all the $F$-terms vanish and
 large supergravity corrections to the $\eta$-parameter do {\it not} appear.

The 
one-loop effective potential for the inflaton field reads $
V_{{\rm 1-loop}} = {g^2 \over 2}\xi^2 \left( 1 + {g^2 \over 16\pi^2} {\rm ln}
{\lambda^2 |S|^2 \over Q^2}\right)$.
The end of inflation is determined either by
the failure of the slow-roll conditions or when $S$ approaches $S_c$. 
COBE imposes the following normalization $
5.3\times 10^{-4} = \frac{V^{3/2}}{V'\Mpl^3}$. T
his gives with the above potential $
\sqrt{\xi_{{\rm COBE}}}=6.6\times 10^{15}\: {\rm GeV}$.
Notice that his normalization is {\it independent} from the gauge coupling
 constant $g$.
The spectral index results $
n=1-\frac{2}{N}=(0.96-0.98)$.

{\it Advantages and disadvantages.}~~
Now, are we happy with such a COBE normalized value of $\xi$? The answer
depends upon the origin of the FI $D$-term. Let us distinguish three
different options:

$\star$~~ {\sl The Ugly}: Since the  FI $D$-term is not forbidden by any
symmetry, it may be  put
{\it by hand} in the theory from the very beginning. If one is ready to take
such step, successful $D$-term inflation scenarios
 in the framework of supersymmetric
GUT's have been already constructed \cite{dvaliriotto}.   

$\star$~~{\sl The Bad}: The  $D$-term is generated  in some  
low-energy effective theory after some degrees of freedom have
 been integrated out. However, to do so, one has
 presumably to break supersymmetry by some $F$-terms 
present in the sector which the heavy fields  belong to and
 to generate the $D$-term by loop corrections. As a result,
  it turns out that    $\langle D\rangle\ll \langle F^2\rangle$, unless
 some fine-tuning is called for, and large supergravity corrections to
 $\eta$ appear again. 

$\star$~~{\sl The Good}:   
 The  Fayet-Iliopoulos term is the one emerging from string theories and the
 $U(1)$  anomaly is cancelled by the Green-Schwarz
mechanism. This is certainly a positive   aspect. 
 Let us discuss some other aspects  of such 
intriguing possibility: {\it 1)} since in realistic models there appear
a plethora of fields and many of them charged under the anomalous $U(1)$, one
has always to check that during inflation there is no  field which
is allowed to take expectation value of the
order of $\sqrt{\xi}$,  thus making the $D$-term vanishing and stopping
inflation. This analysis seems to be missing in the
literature; {\it 2)} in string 
theories the FI $D$-term is  dilaton-dependent and
we do not  know how the dilaton field behaves during the inflationary stage. A
common assumption is that some mechanism stabilizes the dilaton and
that  the latter is  reduced to a  frozen degree of freedom during inflation. 
This assumption might
turn out to be completely wrong; {\it 3)} 
the value of $\sqrt{\xi}$ is expected to 
be of the order of the 
stringy scale,   $(10^{17}-10^{18})$ GeV or so. This value is at least one order
of magnitude higher than the COBE normalized one. How can we solve
this problem? One possible solution to the mismatch between the stringy
and the COBE scales is to go to the strong coupling limit of the heterotic
$E_8\otimes E_8$ string theory compactified on a CY manifold and
 described by the 11-D  
M-theory compactified on CY$\times S^1/Z_2$ \cite{witten}.
 Here the fundamental scale is the
mass scale $M_{11}$  associated to the extra 11th dimension.  This scale 
may be much smaller
than $\Mpl$. However, it is not  clear whether $\Mpl$ will be replaced by
$M_{11}$ in the expression for the FI $D$-term and, if so, whether this will
 be the only change in the
theory; {\it 4)} another point we would like to comment on is the following:
 when the field $\phi_{-}$ in the toy model described above 
rolls down to its 
present day value $\langle\phi_{-}\rangle=\sqrt{\xi}$
  to terminate inflation, 
cosmic strings may be form since the abelian 
gauge group $U(1)$ is broken to unity. 
As it is known, 
stable cosmic strings arise when the manifold ${\cal M}$ 
of degenerate vacua has a non-trivial first homotopy group, 
$\Pi_1({\cal M})\neq {\bf 1}$. 
The fact that at the end of hybrid inflationary
 models the formation of cosmic strings may occur 
was already noticed 
in the context of global supersymmetric theories\cite{j1} 
and  in the context of supergravity 
theories\cite{linderiotto}.
 In $D$-term inflation the string 
per-unit-length is given by $\mu=2\pi\xi$. 
Cosmic strings forming at the end of $D$-term 
inflation are very heavy and temperature 
anisotropies may arise both from the inflationary 
dynamics and from the presence of cosmic strings. 
From recent numerical simulations on the 
cosmic microwave background  anisotropies induced by 
cosmic strings \cite{a1}
it is possible to infer than 
this mixed-perturbation scenario \cite{linderiotto}
leads to the COBE normalized value
 $\sqrt{\xi}=4.7\times 10^{15}$ GeV, 
which is of course smaller  than the value obtained 
in the absence of cosmic strings.  Moreover, cosmic strings 
contribute to the angular spectrum an amount of order of 75\% 
in $D$-term inflation, which might render the 
angular spectrum, when both cosmic strings and inflation 
contributions are summed up, too smooth to be in agreement 
with present day observations  \cite{a1}. 
All these considerations and, above all, the fact that the value of 
  $\sqrt{\xi}$ is further reduced with respect to the case in 
which cosmic strings are not present,  would appear  to   exacerbate 
the problem of reconciling the value of  $\sqrt{\xi}$ suggested by COBE 
with the value inspired by string theories when cosmic strings are present. 
However,  even though cosmic strings are generally produced,  this is not 
always  true and the question is very model-dependent. Indeed,   in string
theory model building there appear often many abelian factors and the 
issue of cosmic string formation must be analyzed case by case. Of course,
the most preferable case in the one in which cosmic strings do not form at 
all. 
In conclusion, an inflationary stage dominated by a 
$D$-term avoids the slow-roll problem of inflation in supergravity 
and can naturally emerge in theories with a non-anomalous
 or anomalous $U(1)$ gauge symmetry. In  the latter  case,
 however, the scale of inflation as
 imposed by  the COBE  normalization is in contrast with the value  fixed 
by the Green-Schwarz mechanism of anomaly cancellation. This makes life even
more
interesting!

\section*{Acknowledgments}
The author would like to thank M. Dine, E. Dudas,  
G. Dvali,  D.H. Lyth, M. Quiros, G.G. Ross
 for many enlightening discussions about different aspects of the
subject. It is also a pleasure to thank the organizers of the COSMO97 
Conference for the stimulating atmosphere  they have been able to
create.


\section*{References}
\begin{thebibliography}{99}





\bibitem{CLLSW} E. J. Copeland, A. R. Liddle, D. H. Lyth, E. D. Stewart
        and D. Wands, Phys. Rev. D {\bf 49}, 6410  (1994).
\bibitem{coughlan} M. Dine, W. Fischler and D. Nemeschansky,
        Phys. Lett. {\bf 136B}, 169 (1984);
G. D. Coughlan, R. Holman, P. Ramond and G. G. Ross,
        Phys. Lett. {\bf 140B}, 44 (1984).

\bibitem{lr} D. H. Lyth and A. Riotto, {\it 
Models of inflation, particle physics 
and the spectral index of the density perturbations}, to 
be submitted to Phys. Rep.

\bibitem{bindvali} P. Binetruy and G. Dvali,  
Phys. Lett. {\bf B388}, 241 (1996).
\bibitem{halyo} E. Halyo, Phys. Lett. {\bf B387}, 43 (1996).

\bibitem{dr} D.H. Lyth and A. Riotto, hep-ph/9707273.

\bibitem{u(1)A} M. Green and J. Schwarz, Phys. Lett. {\bf B149}, 117 (1984).


 \bibitem{fi} M. Dine, N. Seiberg and 
E. Witten, Nucl. Phys. {\bf B289}, 
585 (1987); J. Atick, L. Dixon and A. Sen, Nucl. Phys. {\bf B292},
  109 (1987); M. Dine, I. Ichinose and 
N. Seiberg, Nucl. Phys. {\bf B293},  253 (1987). 

\bibitem{dvaliriotto} G. Dvali and A. Riotto, hep-ph/9706408.  

\bibitem{witten} E. Witten, Nucl. Phys. {\bf B471}, 135 (1996).



\bibitem{j1} R. Jeannerot, Phys. Rev. {\bf D53}, 5426 (1996). 

\bibitem{linderiotto} A. D. Linde 
and A. Riotto, hep-ph/9703209, to be published in Phys. Rev. {\bf D}.  

\bibitem{a1} B. Allen,  R.R. Caldwell, S. Dodelson, 
 L. Knox, E.P.S. Shellard 
and  A. Stebbins, astro-ph/9704160; 
Ue-Li Pen,  U. Seljak and N.  Turok,  astro-ph/9704165. 








\end{thebibliography}
\end{document}